\documentclass{osa-article}

\journal{oe}
\begin{document}

\title{Spectral pre-modulation of training examples enhances the spatial resolution of the Phase Extraction Neural Network (PhENN)}

\author{Shuai Li\authormark{1,*}  and George Barbastathis\authormark{1,2}}

\address{\authormark{1}Department of Mechanical Engineering, Massachusetts Institute of Technology, 77 Massachusetts Avenue, Cambridge, Massachusetts 02139, USA\\
\authormark{2}Singapore-MIT Alliance for Research and Technology (SMART) Centre, One Create Way, Singapore 117543, Singapore}

\email{\authormark{*}shuaili@mit.edu} %% email address is required



\begin{abstract}
The Phase Extraction Neural Network (PhENN) \cite{sinha2017lensless} is a computational architecture, based on deep machine learning, for lens-less quantitative phase retrieval from raw intensity data. PhENN is a deep convolutional neural network trained through examples consisting of pairs of true phase objects and their corresponding intensity diffraction patterns; thereafter, given a test raw intensity pattern PhENN is capable of reconstructing the original phase object robustly, in many cases even for objects outside the database where the training examples were drawn from. Here, we show that the spatial frequency content of the training examples is an important factor limiting PhENN's spatial frequency response. For example, if the training database is relatively sparse in high spatial frequencies, as most natural scenes are, PhENN's ability to resolve fine spatial features in test patterns will be correspondingly limited. To combat this issue, we propose ``flattening'' the power spectral density of the training examples before presenting them to PhENN. For phase objects following the statistics of natural scenes, we demonstrate experimentally that the spectral pre-modulation method enhances the spatial resolution of PhENN by a factor of $2$. 
\end{abstract}

%%%%%%%%%%%%%%%%%%%%%%%%%%  body  %%%%%%%%%%%%%%%%%%%%%%%%%%
\section{Introduction \label{sec:intro}}

The use of machine learning architectures is a relatively new trend in computational imaging and rapidly gaining popularity. Originally it was proposed for imaging through scatter using the older neural network format of support vector machines \cite{horisaki2016learning}. Subsequently, contemporary Deep Neural Networks (DNNs) have been applied successfully to the same problem of imaging through scatter \cite{lyu2017exploit, li2018imaging}, as well as tomography \cite{jin2017deep}, lensless quantitative phase retrieval \cite{sinha2017lensless, rivenson2018phase}, microscopy \cite{rivenson2017deep}, GHOST imaging \cite{lyu2017deep}, imaging through fiber bundles \cite{Borhani:18}, and imaging at extremely low light levels \cite{goy2018low}.  

The main motivation for the use of machine learning is to overcome certain deficits of traditional computational imaging approaches. The latter are based on convex optimization, structured so that the optimal solution is as close as possible to the true object. The functional to be minimized is specified by the physical model  $H$ of the imaging process, also referred to as forward operator; and by prior knowledge $\Phi(f)$ about the class of objects being imaged, also known as regularizer. The inverse (estimate) $\hat{f}$ of an object $f$ is obtained from a measurement $g$ as
\begin{equation}
\hat{f}={\text{argmin}}_{f} \left\{\rule[-1ex]{0cm}{3ex} 
\left|\!\left| Hf-g \right|\!\right|^2 +
\alpha \Phi(f)
\right\}
\label{eq:minfunc}
\end{equation}
The regularization parameter $\alpha$ expresses the imaging system designer's relative belief in the measurement {\it vs.} belief in the available prior knowledge about the object class.

Clearly, the performance of (\ref{eq:minfunc}) in terms of producing acceptable inverses is crucially dependent upon correct and explicit knowledge of both $H$ and $\Phi$, and judicious selection of the parameter $\alpha$ \cite{liao2009selection}. In situations where this knowledge is questionable or not explicitly available, deep machine learning approaches become appealing as an effort to {\em learn} the missing knowledge implicitly through examples. Instead of (\ref{eq:minfunc}), the object estimate is then obtained as
\begin{equation}
\hat{f} = \text{DNN}(g),
\label{eq:dnn}
\end{equation}
where $\text{DNN}(.)$ denotes the output of the trained deep neural network. 

Notation (\ref{eq:dnn}) may be used for other, non-deep machine learning structures even though they are generally less effective. However, strictly applied, (\ref{eq:dnn}) is limited to the special ``end-to-end'' design where the measurement $g$ from the camera is fed directly to the DNN. In some cases $g$ first goes through a physical pre-processor, and the pre-processor's output is fed into the DNN \cite{rivenson2018phase, goy2018low}; whereas in other cases $g$ is fed multiple times into a cascade of generator DNNs \cite{mardani2017recurrent} to assess the outputs at each step. Developing notation for and carrying out a full debate on the relative merits of these different approaches is beyond the scope of the present paper, where, in any case, we used the end-to-end method (\ref{eq:dnn}) only. 

Just as the performance of minimization principle (\ref{eq:minfunc}) depends upon knowledge of the operators $H$ and $\Phi$,  performance of the DNN principle (\ref{eq:dnn}) depends on the specific DNN architecture chosen (number of layers, connectivity, etc.) and the quality of the training examples. It is the latter aspect of DNN design that we focus on in the present paper. More specifically, we are concerned with the spatial resolution that the DNN can achieve, depending on the spatial frequency content of the examples the DNN is trained with. 

We chose to study this question in the specific context of quantitative phase retrieval. This is a classical problem in optical imaging,  because by virtue of its challenge it evokes elegant solutions and also because it has important applications in biology, medicine, and inspection for manufacturing and security. Traditional approaches include digital holography (DH) \cite{goodman1967digital, rivenson2010compressive, milgram2002computational, brady2009compressive, williams2013digital} and the related phase shifting interferometry method \cite{creath1985phase}, propagation based methods such as the Transport of Intensity Equation (TIE) \cite{teague1983deterministic, kou2010transport, paganin1998noninterferometric, schmalz2011phase, waller2010phase, waller2011phase, tian2013compressive, pan2014contrast, zhu2014low} and iterative methods such as the Gerchberg-Saxton-Fienup algorithm \cite{gerchberg1972practical, fienup1978reconstruction, gonsalves1976phase, fienup1986phase, bauschke2002phase}. 

The end-to-end residual convolutional DNN solution to lens-less quantitative phase retrieval is PhENN \cite{sinha2017lensless}, shown to be robust to errors in propagation distance and fairly well able to generalize to test objects from outside the databases used for training. In the present paper, we implemented PhENN in a slightly different optical hardware configuration, described in Section~\ref{sec:archi-opt}. The computational architecture, described in Section~\ref{sec:archi-dnn}, was similar to the original PhENN except here we used the Negative Pearson Correlation Coefficient (NPCC) as training loss function. This has a small beneficial effect in the reconstructions, but necessitates a histogram calibration procedure, described in Section~\ref{sec:cali}, to remove linear amplification and bias in the reconstructed phase images.

From the point of view of the original inverse problem formulation (\ref{eq:minfunc}), PhENN in effect has to learn both the forward operator $H$ and the prior $\Phi$ at the entire range of spatial frequencies of interest. The examples presented to PhENN during training establish the spatial frequency content that is stored in the network weights contributing to the retrieval operation (\ref{eq:dnn}). In principle, this should be sufficient because, if the training examples are representative enough of the object class, then retrieval of each spatial frequency should be learnt proportionally to that spatial frequency's presence in the database. In practice, however, we found that spatial frequencies with relatively low representation in the database tend to be overshadowed by the more popular spatial frequencies, perhaps due to the nonlinearities in the network training process and operation. 

Invariably, high spatial frequencies tend to be less popular in most available databases. ImageNet, in particular, exhibits the well-known inverse-square power spectral density of natural images, as we verify in Fig.~\ref{fig:spec}. This means that high spatial frequencies are inherently under-represented in PhENN training. Compounded by the nonlinear suppression of the less popular spatial frequencies due to PhENN nonlinearities, as mentioned above, this results in low-pass filtering of the estimates and loss of fine detail. Detailed analysis of this effect is presented in Section~\ref{sec:train}.

To better recover high spatial frequencies in natural objects then, one should emphasize high spatial frequencies more during training; this may be achieved, for example, by flattening the power spectral density of the training examples {\em before} they are presented to the neural network. It would appear that this spectral intervention violates the object class priors: PhENN does not learn the priors of ImageNet itself, it rather learns an edge-enhanced version of the priors. Yet, in practice, again probably because of nonlinear PhENN behavior, we found this spectral pre-modulation strategy to work quite well. The detailed approach and results are found in Section~\ref{sec:res-enh}.
  
It is worth mentioning here that the first, to our knowledge, explicit experimental analysis of a DNN's spatial resolution was conducted on IDiffNet in the context of imaging through diffuse media \cite{li2018imaging}. We chose to pursue the issue further in the present paper but on a different optical problem because spatial resolution in quantitative phase retrieval, in addition to also being worthwhile, is not impacted by the extreme ill-posedness of diffuse media. Even though we have not tried extensively beyond phase retrieval, pre-processing of training examples by spectral manipulation might have merit for several other challenging imaging problems. 

\section{Imaging system architecture \label{sec:archi}}

\subsection{Optical configuration \label{sec:archi-opt}}

\begin{figure}[h!]
\centering\includegraphics[width=0.7\linewidth]{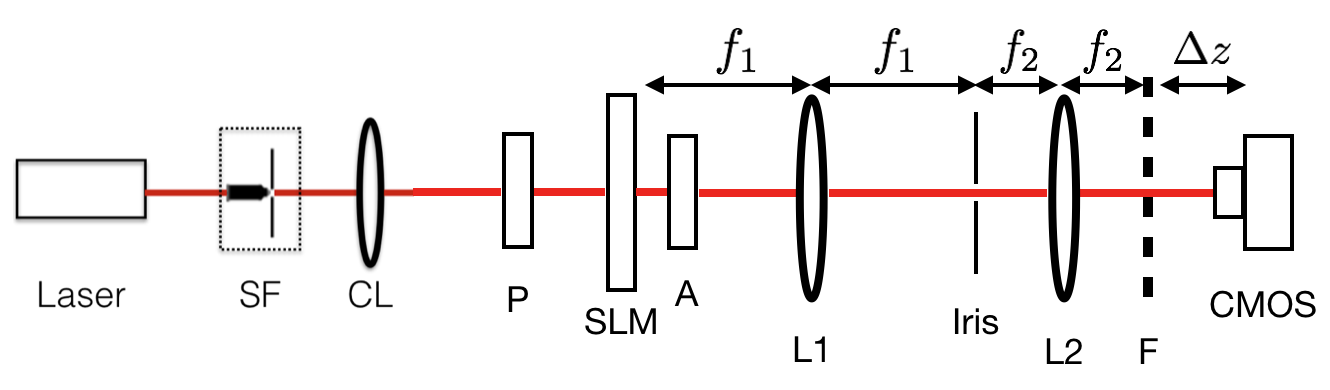}
\caption{Optical configuration. SF: spatial filter; CL: collimating lens; P: linear polarizer; A: analyzer; SLM: spatial light modulator; L1 and L2: plano-convex lenses; F: focal plane of L2}
\label{fig:setup}
\end{figure}

Our optical configuration is shown in Fig. \ref{fig:setup}. Unlike \cite{sinha2017lensless}, a transmissive spatial light modulator (SLM) (Holoeye, LC2012, pixel size $36\mu\text{m}$) is used in this system as a programmable phase object $f$ representing the ground truth. The transmissive SLM is coherently illuminated by a He-Ne laser light source (Research Electro-Optics, Model 30995, $633$nm). The  light is transmitted through a spatial filter consisting of a microscope objective (Newport, M-60X, $0.85$NA) and a pinhole aperture ($D = 5\mu\text{m}$) and then collimated by a lens (focal length $200\text{mm}$) before illuminating the SLM. A telescope consisting of two plano-convex lenses $L_{1}$ and $L_{2}$ is placed between the SLM and a CMOS camera (Basler, A504k, pixel size $12\mu\text{m}$). The CMOS camera captures the intensity $g$ of the diffraction pattern produced by the SLM at a defocus distance $\Delta z=50\text{mm}$. The focal lengths of $L_{1}$ and $L_{2}$ are set to $f_{1}=150\text{mm}$ and $f_{2}=50\text{mm}$, respectively. As a result, this telescope demagnifies the object by a factor of $3$, consistent with the ratio between SLM and CMOS camera pixel sizes. An iris with diameter $5\text{mm}$ is placed at the pupil plane of the telescope to keep the $0^{\text{th}}$ diffracted order of the SLM and filter out all the other orders. 

The modulation performance of the SLM depends on the input and output polarizations, which are controlled by the polarizer $P$ and the analyzer $A$, respectively.  In order to realize phase-mostly modulation, we set the incident beam to be linearly polarized at $310^{\circ}$ with respect to the vertical direction and also set the analyzer to be oriented at $5^{\circ}$ with respect to the vertical direction. The specific calibration curves for the SLM's modulation performance can be found in \cite{li2018quantitative}. In the present paper, all the training and testing objects are of size $256\times 256$.  They are zero-padded to the size $1024\times768$, before being uploaded to the SLM.  For the diffraction patterns captured by the CMOS camera, we crop the central $256\times256$ region for processing.

\subsection{Neural network architecture and training \label{sec:archi-dnn}}
\begin{figure}[h!]
\centering\includegraphics[width=\linewidth]{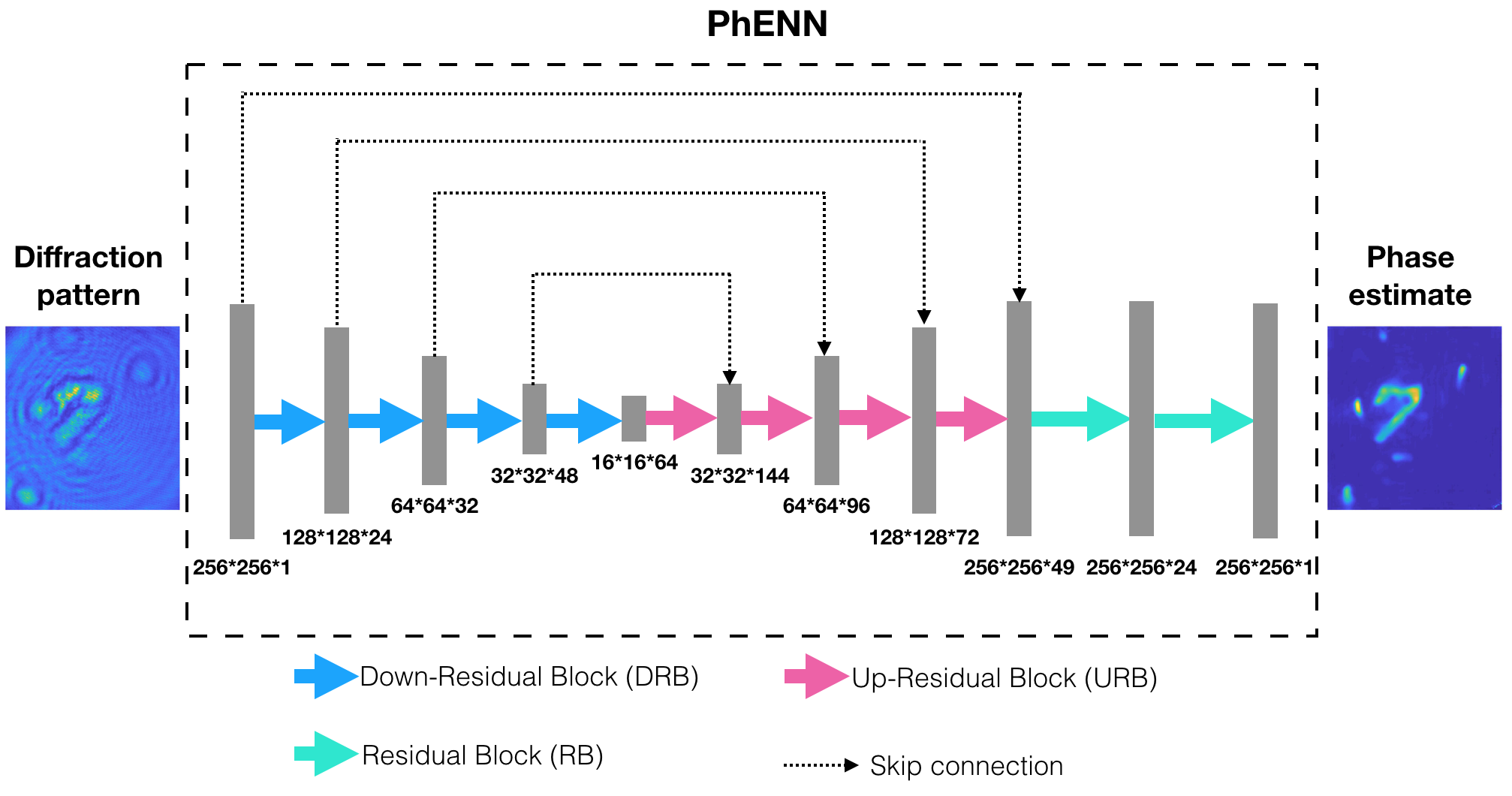}
\caption{Phase extraction neural network (PhENN) architecture}
\label{fig:PhENN}
\end{figure}

Similar to \cite{sinha2017lensless}, the phase extraction neural network (PhENN) that we implement in this paper follows the U-net architecture \cite{ronneberger2015u} and utilizes residuals to facilitate learning (ResNet \cite{he2016deep}.) The detailed architecture is shown in Fig. \ref{fig:PhENN}. PhENN input is the intensity $g$, and successively passes through $4$ down-residual blocks (DRBs) for feature extraction. The extracted feature map then successively passes through $4$ up-residual blocks (URBs) and $2$ residual blocks (RBs) for pixel-wise regression and at the last layer outputs the estimate $\hat{f}$ of the object phase. Skip connections are used in the architecture to pass downstream local spatial information learnt in the initial layers. More details about the structures of the DRBs, URBs and RBs can be found in \cite{sinha2017lensless}.

Unlike \cite{sinha2017lensless}, here we use the Negative Pearson Correlation Coefficient (NPCC) as loss function \cite{li2018imaging} to train PhENN. The NPCC loss function is defined as
\begin{equation}
{\cal L}=\sum_k {\cal E}_{\text{NPCC}}\left(f_k, \hat{f}_k\right), \qquad \text{where}
\end{equation}
\begin{equation}
{\cal E}_{\text{NPCC}}\left(f_k, \hat{f}_k\right)\equiv
(-1)\times\frac{\sum_{i,j}^{}\left(\rule[-1ex]{0cm}{3ex} f_k(i,j)-\left<f_k\right>\right)
\left(\rule[-1ex]{0cm}{3ex} \hat{f}_k(i,j)-\left<\hat{f}_k\right>\right)}
{\sqrt{\sum_{i,j}\left(\rule[-1ex]{0cm}{3ex} f(i,j)-\left<f_k\right>\right)^2}
\sqrt{\sum_{i,j}\left(\rule[-1ex]{0cm}{3ex} \hat{f}(i,j)-\left<\hat{f}_k\right>\right)^2}};
\label{eq:npcc}
\end{equation}
$f$ and $\hat{f}$ are the true object and the object estimate according to (\ref{eq:dnn}), respectively; the summations take place over all pixels $(i,j)$ and training example labels $k$; and $\left<.\right>$ denotes spatial averaging. We have found the NPCC to generally result in better DNN training in the problems that we examined, especially for objects that are spatially sparse \cite{li2018imaging}. However, some care needs to be taken when the estimate $\hat{f}$ is not affine-invariant; we discuss this immediately below.

\subsection{Calibration of PhENN output trained with NPCC} \label{sec:cali}
From the definition (\ref{eq:npcc}) it follows that for any function $\psi$ and arbitrary real constants $a$ and $b$ representing linear amplification and bias, respectively,  
\[
{\cal E}_{\text{NPCC}}(\psi, a\psi+b)=-1.
\] 
In other words, a DNN trained with NPCC as loss function can only produce affine transformed estimates; there is no way to enforce the requirement $a=1$, $b=0$ which would guarantee linear amplification- and bias-free reconstruction and is especially important for quantitative phase imaging. Neither does there exist a way that we know of to predetermine the values of $a$ and $b$ through specific choices in DNN training. 

\begin{figure}[h!]
\centering\includegraphics[width=0.6\linewidth]{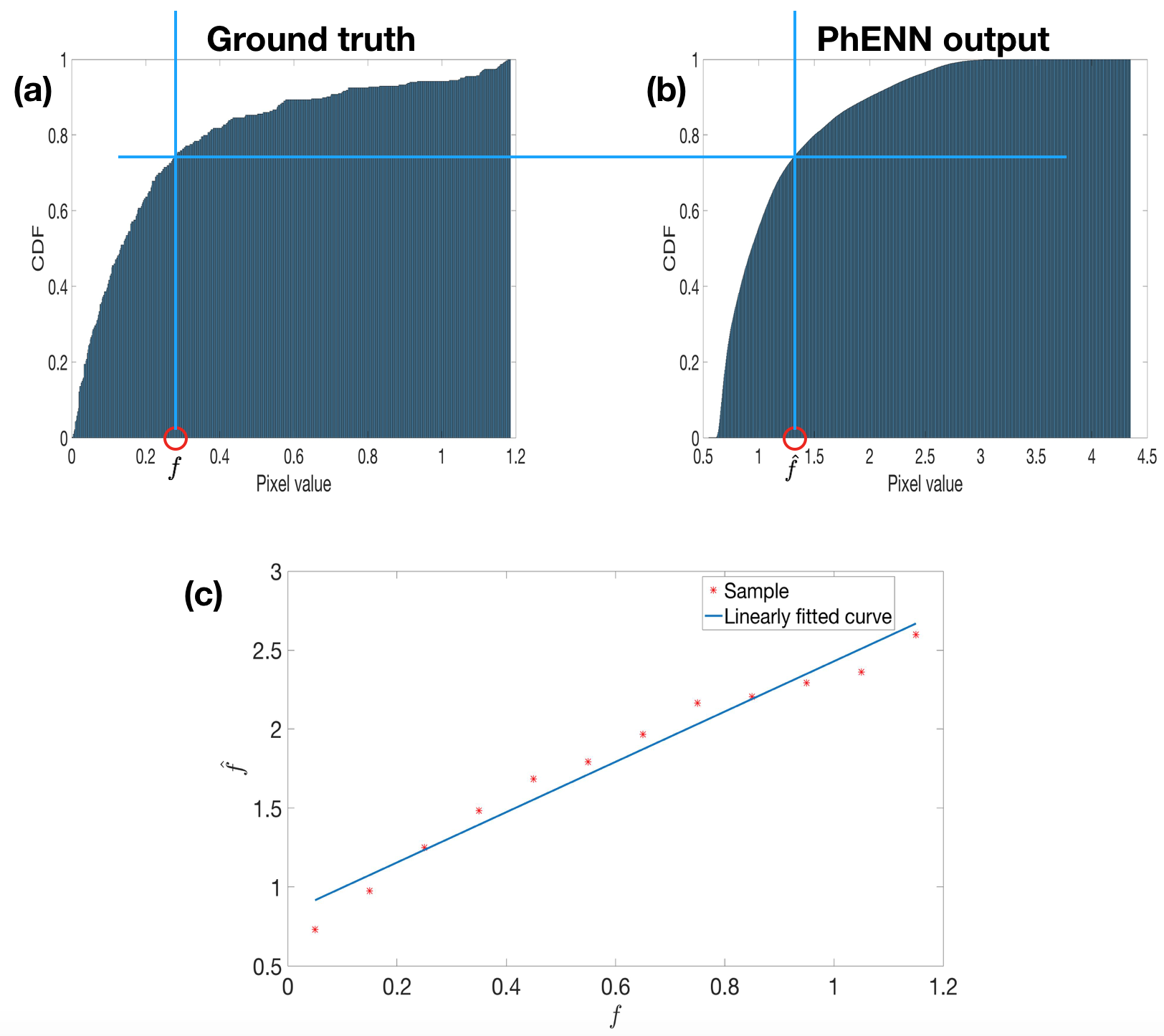}
\caption{Calibration process. (a) Cumulative distribution function (CDF) of the ground truth. (b) Cumulative distribution function (CDF) of the PhENN output. (c) Linear curve fitting.}
\label{fig:cal}
\end{figure}

Therefore, after DNN training a calibration step is required to determine the values of $a$ and $b$ that have resulted so that they can be compensated. This is realized by histogram matching according to the process shown in Fig.~\ref{fig:cal}. Given a set of calibration data,  we compute the cumulative distribution functions (CDFs) for the ground truth values as well as the PhENN output values, as shown in Fig.~\ref{fig:cal} (a) and (b). For an arbitrary value $f$ in the ground truth, we find its corresponding PhENN output value $\hat{f}$ that is at the same CDF level; and repeat the process for several $(f, \hat{f})$ samples. Subsequently, the values of $a$ and $b$ are determined by linear fitting of the form $\hat{f}=af+b$, as shown in Fig. \ref{fig:cal}(c). 

\section{Resolution analysis of ImageNet-trained PhENN \label{sec:train}}

In \cite{sinha2017lensless}, we trained separate PhENNs using the databases Faces-LFW \cite{huang2007labeled} and ImageNet \cite{russakovsky2015imagenet} and found that both PhENNs generalize to test objects both within and outside these two databases. In the present paper, we restrict our analysis to the ImageNet database only. 

In the PhENN training phase, a total of $10,000$ images selected from the ImageNet database are uploaded to the SLM and the respective diffraction patterns are captured by the CMOS.  For testing, we use a total of $471$ images selected from several different databases: $50$ Characters, $40$ Faces-ATT \cite{samaria1994parameterisation}, $60$ CIFAR \cite{krizhevsky2009learning}, $100$ MNIST \cite{lecun2010mnist}, $100$ Faces-LFW, $100$ ImageNet,  $20$ resolution test patterns \cite{li2018imaging}, and $1$ all-zero (dark) image. The diffraction pattern corresponding to the all-zero image is used as the background. For every test diffraction pattern that we capture, we first subtract the background and then normalize, before feeding into the neural network.

\subsection{Reconstruction results}

\begin{figure}[h!]
\centering\includegraphics[width=0.7\linewidth]{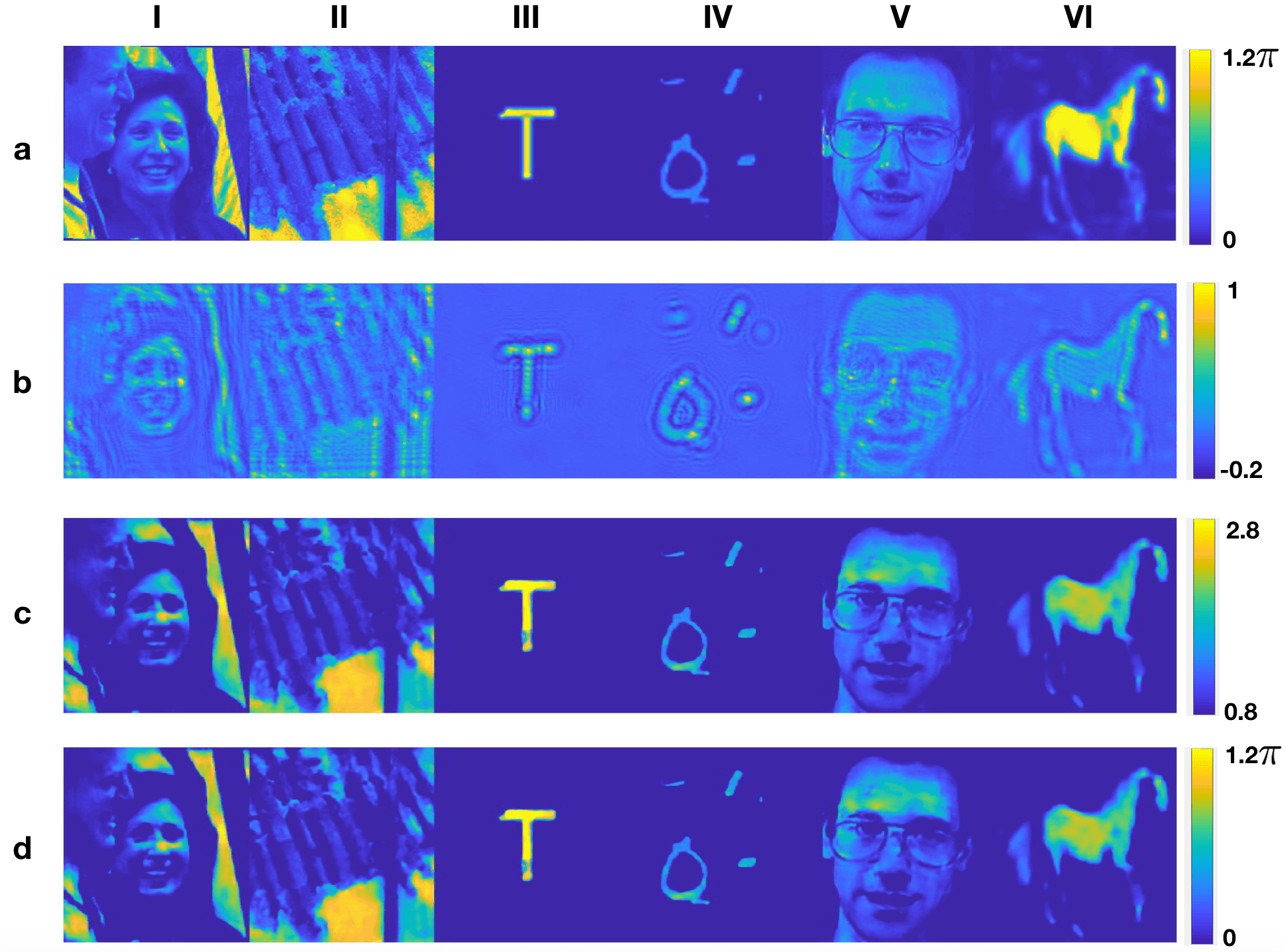}
\caption{Reconstruction results of PhENN trained with ImageNet. (a) Ground truth for the phase objects. (b) Diffraction patterns captured by the CMOS (after background subtraction and normalization). (c) PhENN output. (d) PhENN reconstruction after the calibration shown in Section \ref{sec:cali}. Columns (i-vi) correspond to the dataset from which the test image is drawn: (i) Faces-LFW \cite{huang2007labeled}, (ii) ImageNet \cite{russakovsky2015imagenet}, (iii) Characters, (iv) MNIST Digits \cite{lecun2010mnist}, (v) Faces-ATT \cite{samaria1994parameterisation}, or (vi) CIFAR \cite{krizhevsky2009learning}, respectively.}
\label{fig:qual}
\end{figure}

The phase reconstruction results are shown in Fig. \ref{fig:qual}. Here, we use $100$ ImageNet test images as  calibration data to compensate for the unknown affine transform effected by the NPCC-trained PhENN (Section~\ref{sec:cali}).  As expected, PhENN is not only able to quantitatively reconstruct the phase objects within the same category as its training database (ImageNet), but also able to retrieve the phase for those test objects from other databases. This indicates that PhENN has indeed learned a model of the underlying physics of the imaging system or at the very least a generalizable mapping of low-level textures between the phase objects and their respective diffraction patterns.

\subsection{Resolution test \label{sec:res}}
\begin{figure}[h!]
\centering\includegraphics[width=0.7\linewidth]{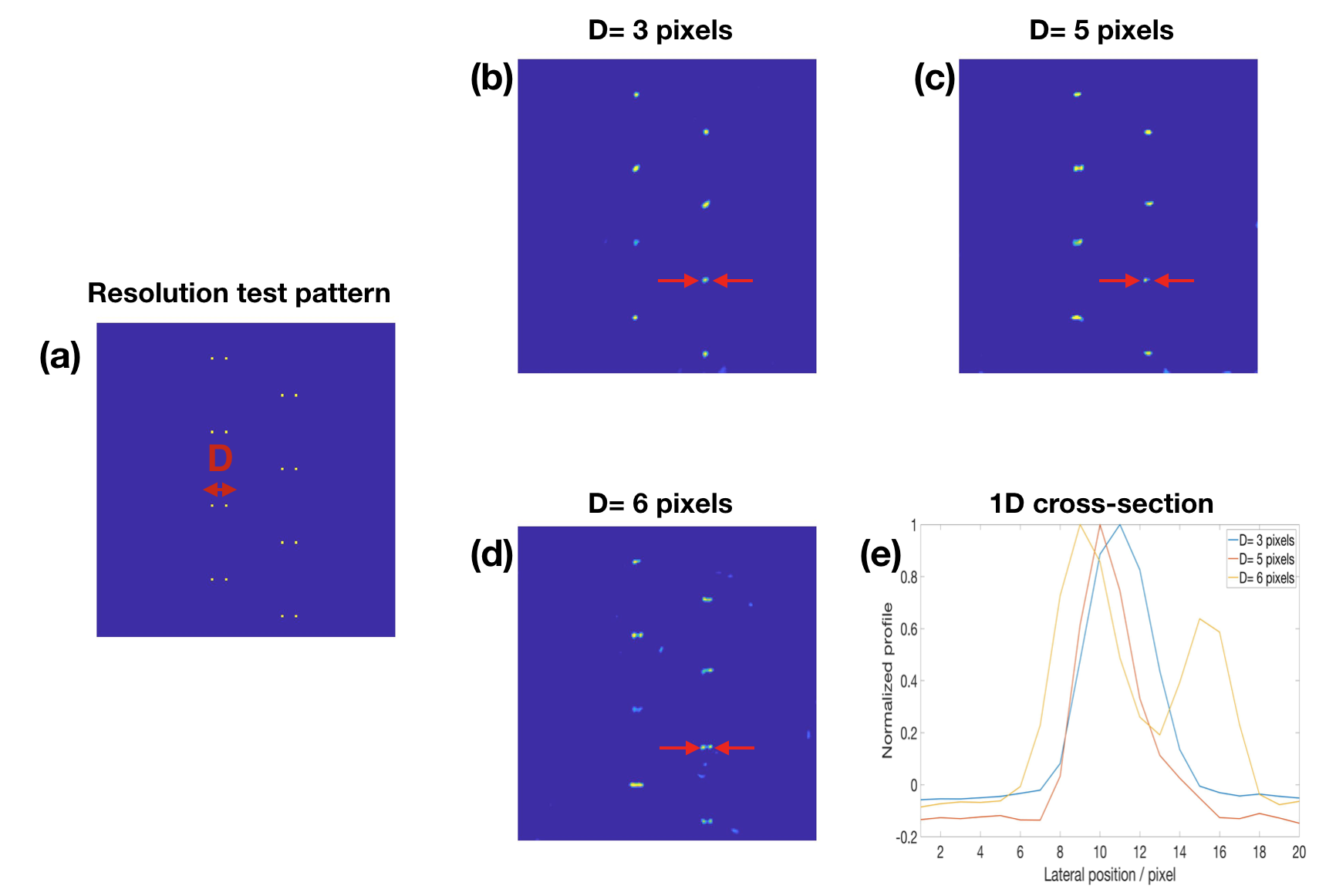}
\caption{Resolution test for PhENN trained with ImageNet. (a) Dot pattern for resolution test. (b) PhENN reconstructions for dot pattern with $D=3$ pixels. (c) PhENN reconstructions for dot pattern with $D=5$ pixels. (d) PhENN reconstructions for dot pattern with $D=6$ pixels. (e) 1D cross-sections along the lines indicated by red arrows in (b)-(d).}
\label{fig:res}
\end{figure}

In order to test the spatial resolution our trained PhENN, we use dot patterns as test objects \cite{li2018imaging}, shown in Fig. \ref{fig:res} (a). Altogether $20$ dot patterns are tested, with  spacing $D$ between dots gradually increasing from $2$ pixels to $21$ pixels. From the resolution test results shown in Fig. \ref{fig:res} it can be observed that the PhENN trained with ImageNet is able to resolve two dots down to $D=6$ pixels but fails to distinguish two dots with spacing $D\leq5$ pixels. Thus, $D\approx 6$ pixels can be considered as the Rayleigh resolution limit of this PhENN for point-like phase objects. 

\section{Resolution enhancement by spectral pre-modulation \label{sec:res-enh}}

In our imaging system, the SLM pixel size limits the spatial resolution of the trained PhENN since the minimum sampling distance in all the training and testing objects displayed on the SLM equals one pixel $d_{p}=36\mu\text{m}$, or maximum spatial frequency $13.9\text{mm}^{-1}$. \footnote{L1 determines the aperture stop with diameter $25.4\text{mm}$, {\it i.e.} a numerical aperture NA$=12.7/150=0.0847$. The nominal diffraction-limited resolution should be $d_{0}=\lambda/(2\text{NA})=3.74\mu\text{m}$. That calculation is irrelevant to PhENN, since objets of that spatial frequency are never presented to it during training.} However, as we saw in Section \ref{sec:res}, the resolution achieved by our PhENN trained with ImageNet database is merely $6$ pixels ($216\mu\text{m}$), much worse than the theoretical value.

The additional factor limiting the spatial resolution of the trained PhENN is the spatial frequency content of the training database. Generally, databases of natural objects, such as natural images, faces, hand-written characters, etc. do not cover the entire spectrum up to $1/(2 d_{0})$. For example, below we analyze the ImageNet database and show that it is dominated by low spatial frequency components, with the prevalence of higher spatial frequencies decreasing quadratically. 

During training, the neural network learns the particular prevalence of spatial frequencies in the training examples as prior $\Phi$, in addition to learning the physical forward operator $H$. What this implies is that the less prevalent spatial frequencies are actually learnt {\em against}, meaning that by presenting them less frequently we may be teaching PhENN to suppress or ignore them. In the rest of this section, we present evidence to corroborate this fact, and suggest as solution a pre-processing step that edge enhances the training examples as a way to impress their importance better upon PhENN. 

\subsection{Spectral pre-modulation}
\begin{figure}[h!]
\centering\includegraphics[width=0.8\linewidth]{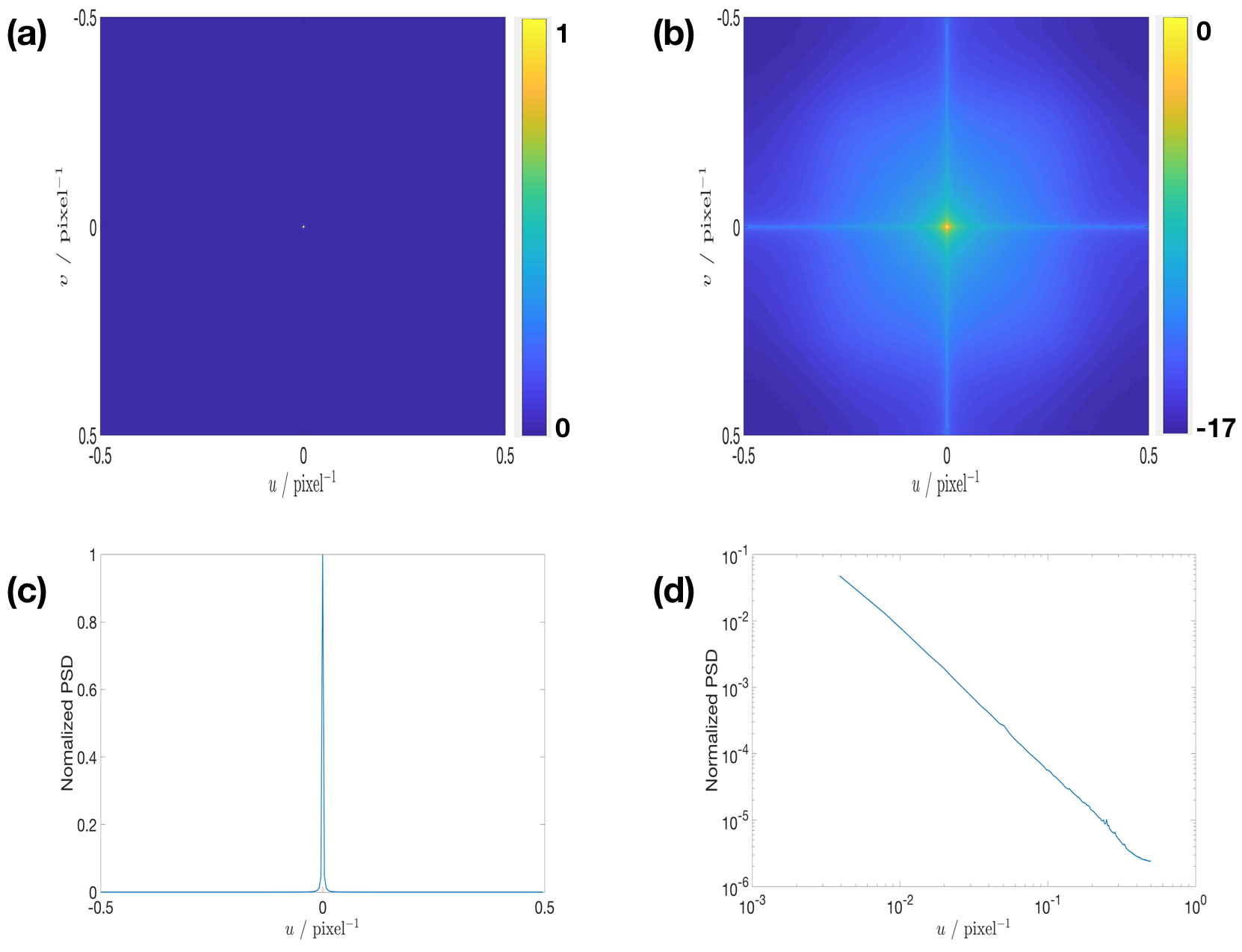}
\caption{Spectral analysis of the ImageNet database. (a\& b) 2D normalized power spectral density (PSD) of the ImageNet database in linear and logarithmic scale. (c\& d) 1D cross-sections along the spatial frequency $u$ of (a\& b), respectively.}
\label{fig:spec}
\end{figure}

The 2D power spectral density (PSD) $S(u,v)$ for the $10,000$ images in the ImageNet is shown in Fig. \ref{fig:spec} (a \&b) in linear and logarithmic scales, respectively;  and in cross-section along the spatial frequency $u$ in Fig. \ref{fig:spec} (c\& d). Not surprisingly \cite{van1996modelling}, the cross-sectional power spectral density follows a power law of the form $\left|u\right|^p$ with $p\approx-2$.

\begin{figure}[h!]
\centering\includegraphics[width=0.7\linewidth]{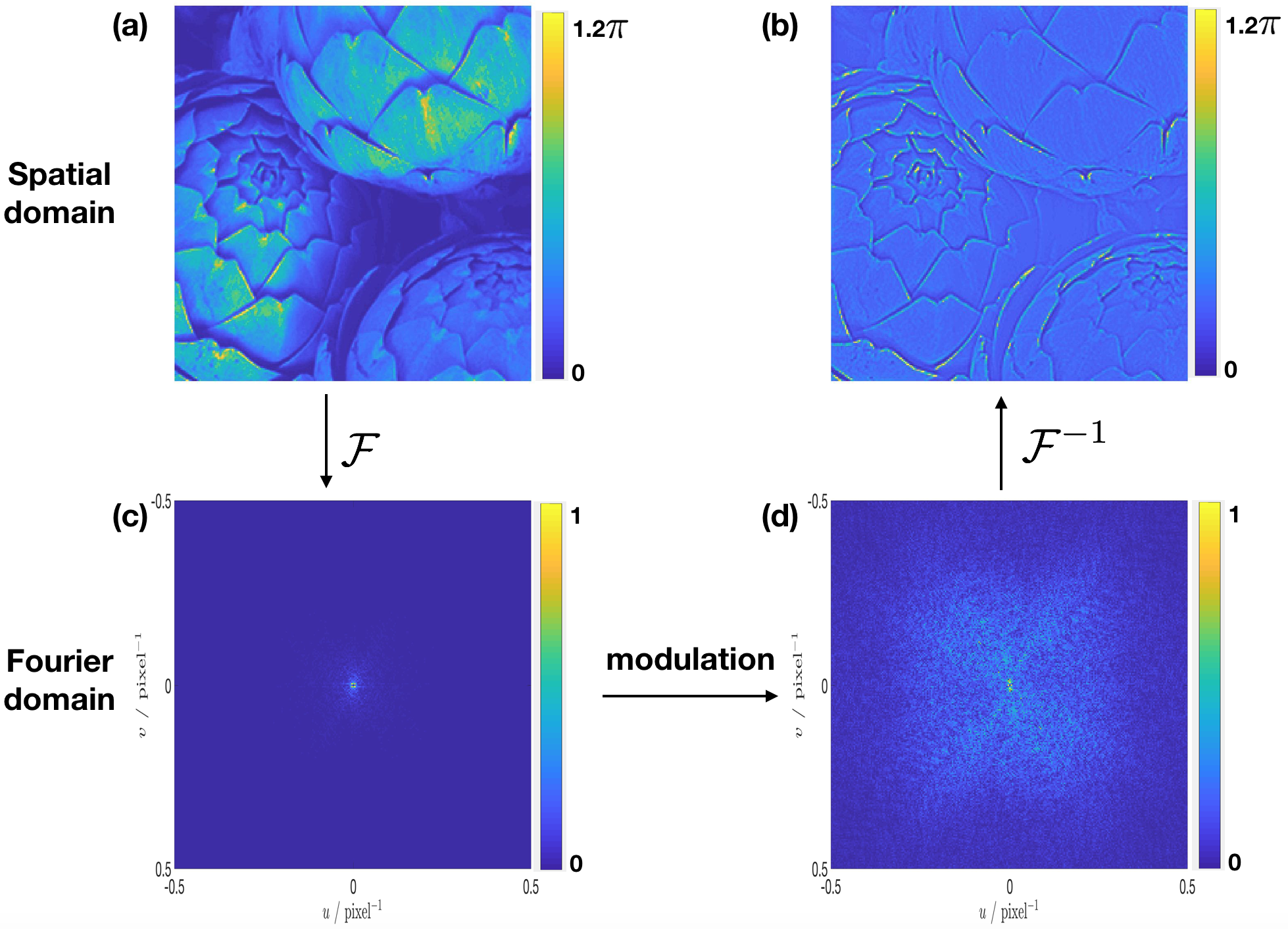}
\caption{Spectral pre-modulation. (a) Original image \cite{russakovsky2015imagenet}. (b) Modulated image. (c) Fourier spectrum of the original image.  (d) Fourier spectrum of the modulated image.}
\label{fig:fsm}
\end{figure}

Therefore, we may approximately represent the 2D PSD of ImageNet database as
\begin{equation}
S(u,v)\propto\left(\sqrt{u^2+v^2}\right)^{-2}=\frac{1}{u^2+v^2}.
\end{equation}
This is flattened by the inverse filter
\begin{equation}
G(u,v)=\sqrt{u^2+v^2}.
\label{eq:G-edge}
\end{equation}
As expected, the high spatial frequency components in the image are amplified after the modulation, as can be seen, for example, in Fig. \ref{fig:fsm}.

\subsection{Resolution enhancement}
\begin{figure}[h!]
\centering\includegraphics[width=0.7\linewidth]{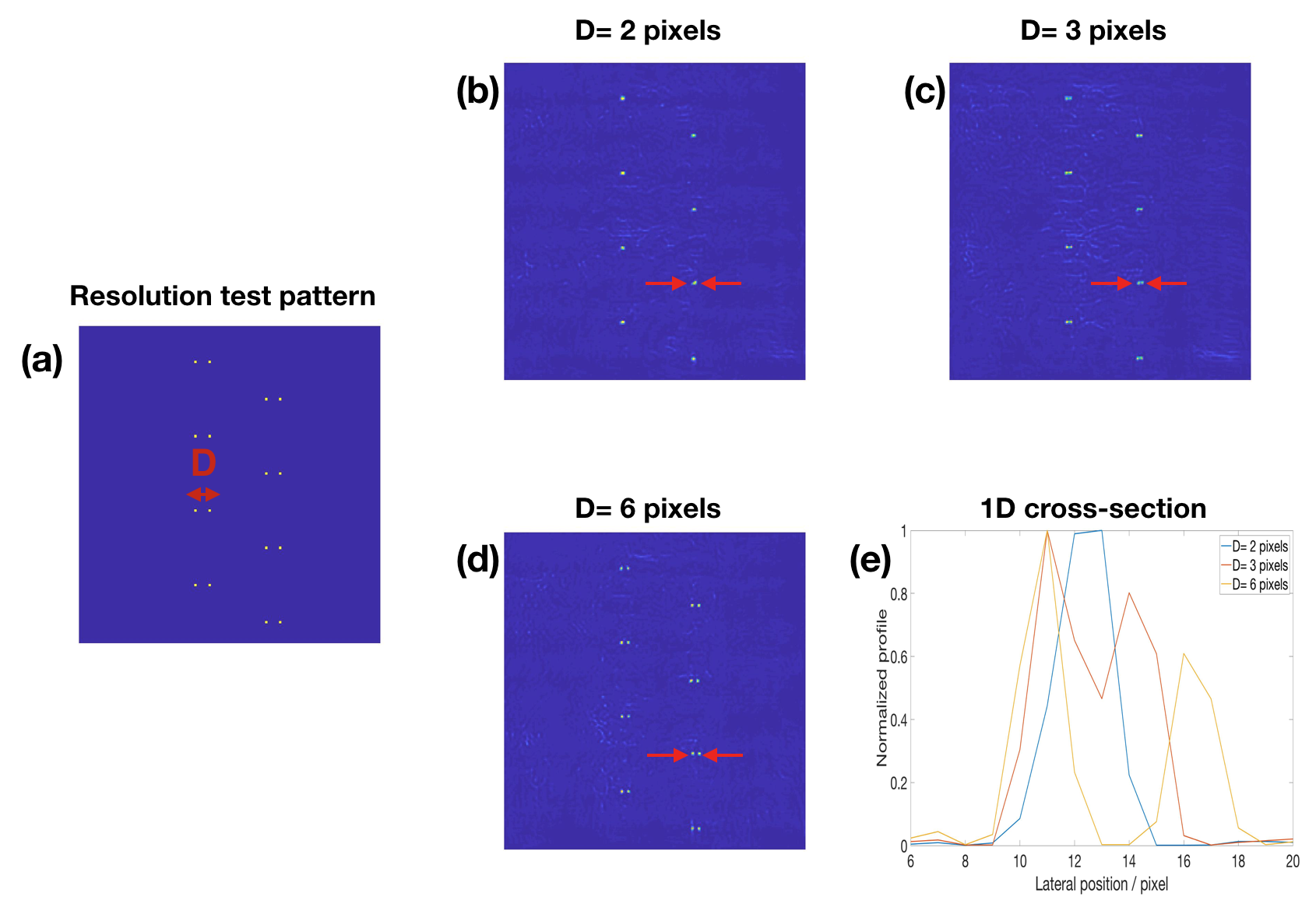}
\caption{Resolution test for PhENN trained with examples from the ImageNet database with spectral pre-modulation according to (\protect{\ref{eq:flatten}}). (a) Dot pattern for resolution test. (b) PhENN reconstructions for dot pattern with $D=2$ pixels. (c) PhENN reconstructions for dot pattern with $D=3$ pixels. (d) PhENN reconstructions for dot pattern with $D=6$ pixels. (e) 1D cross-sections along the lines indicated by red arrows in (b)-(d).}
\label{fig:res2}
\end{figure}

\begin{figure}[h!]
\centering\includegraphics[width=0.7\linewidth]{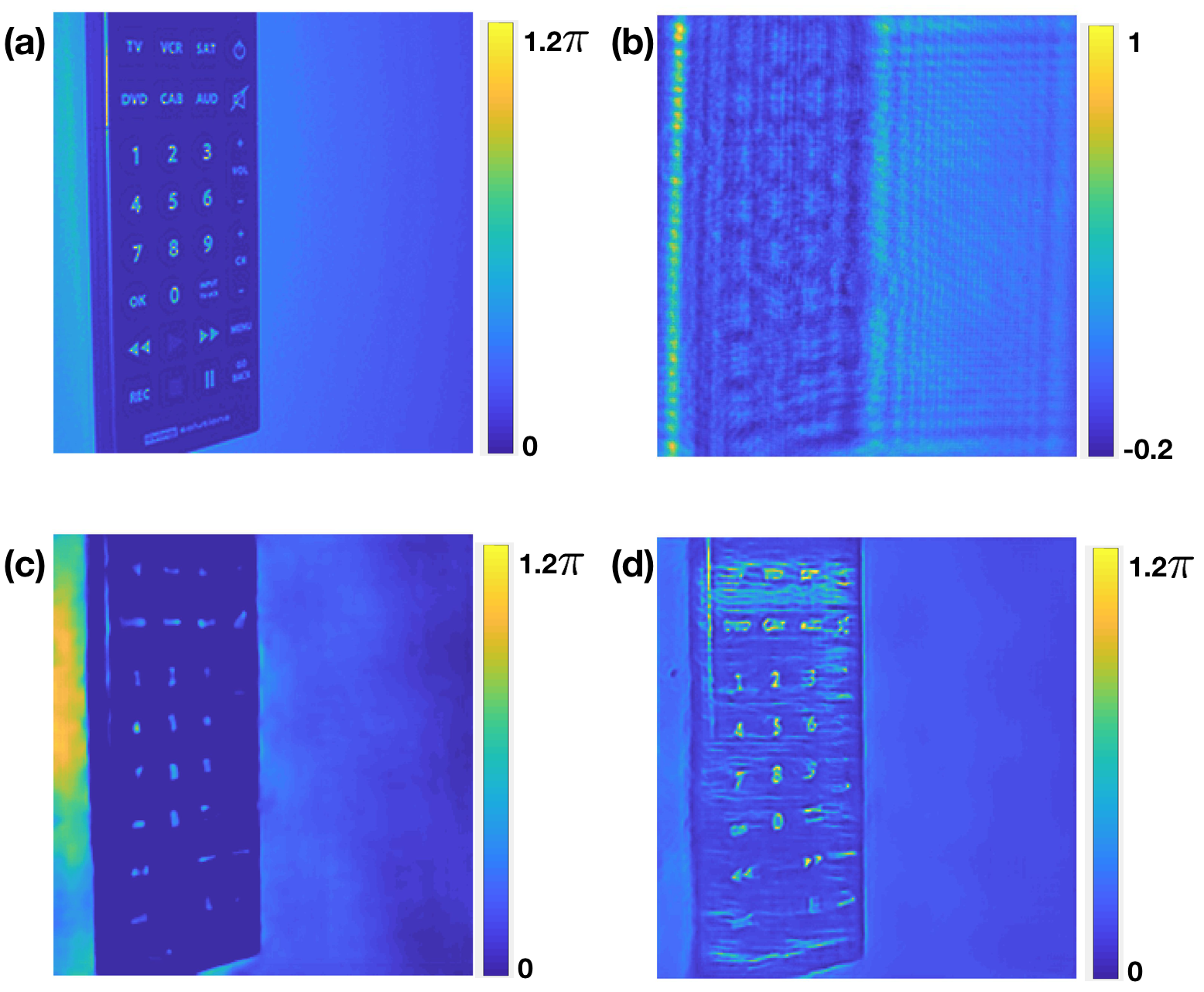}
\caption{Resolution enhancement demonstration. (a) Ground truth for a phase object \cite{russakovsky2015imagenet}. (b) Diffraction pattern captured by the CMOS (after background subtraction and normalization). (c) Phase reconstruction by PhENN trained with ImageNet examples. (d) Phase reconstruction by PhENN trained with ImageNet examples that were spectrally pre-modulated according to (\protect{\ref{eq:flatten}}).}
\label{fig:demo}
\end{figure}

We trained a new PhENN using training examples that were spectrally pre-modulated according to (\ref{eq:G-edge}). That is, we replaced every training example $f(i,j)$ with $f_{\text{e}}(i,j)$, where
\begin{equation}
F_{\text{e}}(u,v)=G(u,v)F(u,v)
\label{eq:flatten}
\end{equation}
and $F$, $F_{\text{e}}$ are the Fourier transforms of $f$, $f_{\text{e}}$, respectively. We also collected the corresponding diffraction patterns $g_{e}(i,j)$. The test examples were left without modulation, {\it i.e.} the same as in the original use of PhENN described in Section \ref{sec:train}. All the training parameters were also kept the same. Both dot pattern and ImageNet test images were used to demonstrate the resolution enhancement, shown in Fig. \ref{fig:res2} and \ref{fig:demo}, respectively.

From Fig. \ref{fig:res2}, we find that with spectral pre-modulation of the training examples according to (\ref{eq:flatten}), PhENN is able to resolve two dots with spacing $D=3$ pixels. Compared with the resolution test results shown in Fig. \ref{fig:res}, it can be said that the spatial resolution of PhENN has been enhanced by a factor of $2$ with the spectral pre-modulation technique. In Fig. \ref{fig:demo}, for the same test image selected from ImageNet database, more details are recovered by the PhENN that was trained with spectrally pre-modulated ImageNet, albeit at the cost of amplifying some noisy features of the object, near edges most notably. 

\begin{figure}[h!]
\centering\includegraphics[width=0.7\linewidth]{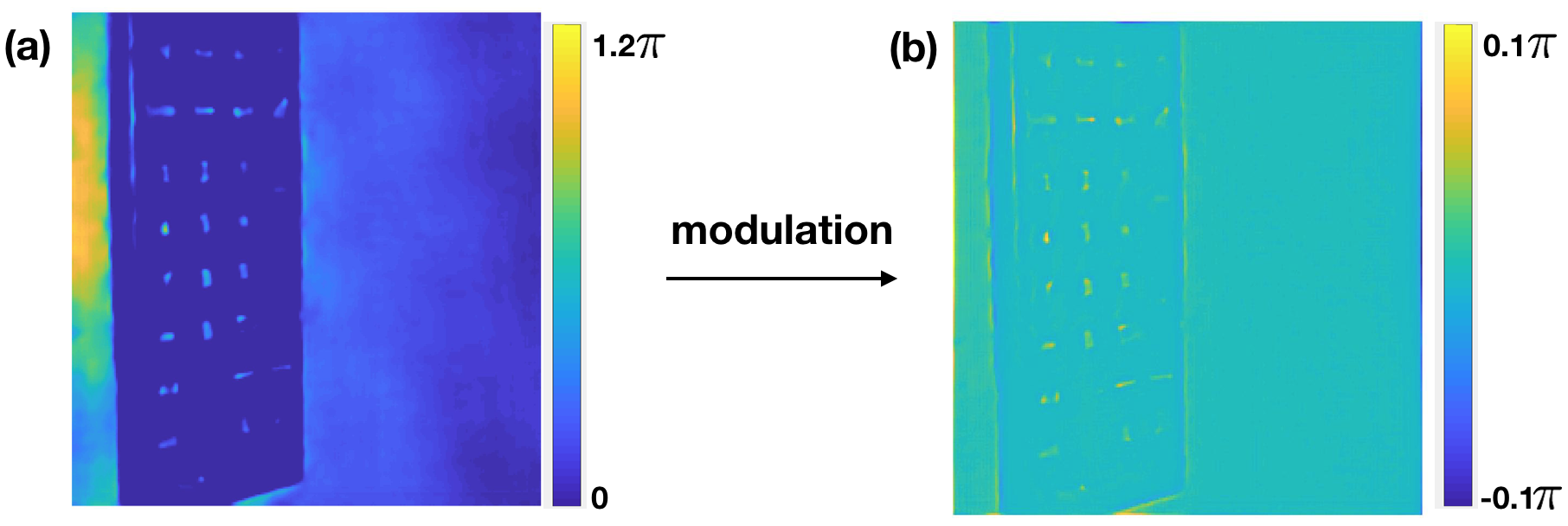}
\caption{Spectral post-modulation. (a) Output of PhENN trained with ImageNet. The same as Fig. \ref{fig:demo} (c). (b) Modulated output.}
\label{fig:comp}
\end{figure}

We also investigated the effect of spectral post-modulation in the original PhENN; that is, if we use a PhENN trained without spectral pre-modulation, and modulate the PhENN output $\hat{f}(i,j)$ according to 
\begin{equation}
\hat{F}_{\text{e}}(u,v)=G(u,v)\hat{F}(u,v)
\label{eq:post-flatten}
\end{equation}
and $\hat{F}$, $\hat{F}_{\text{e}}$ are the Fourier transforms of $\hat{f}$, $\hat{f}_{\text{e}}$, respectively, do we obtain a similar resolution enhancement? The answer is no, as can be clearly verified from the results of Fig. \ref{fig:comp}. 

This negative result illustrates that in the original training scheme (without spectral pre-modulation) the fine details are indeed lost and not recoverable by simple means, e.g. linear post-processing. It also highlights the effect of the nonlinearity in PhENN's operation and bolsters our claim that spectral pre-modulation indeed does something non-trivial: it teaches PhENN a prior, namely how to recover high spatial frequency content. 

\section{Conclusions}
The spectral flattening approach (\ref{eq:flatten}) as pre-modulation is a simple approach that we found to be effective in enhancing PhENN's resolution by a factor of $2$ when trained and tested on ImageNet examples. We have not investigated the performance of other (non-flattening) filters; indeed, it would be an interesting theoretical question to ask: given a particular form of the PSD in the training examples, what is the optimal spectral pre-modulation for improving spatial resolution? 

It is also worth repeating the concern about the priors that PhENN is learning from the spatially pre-modulated examples, that we pointed out in Section~\ref{sec:intro}. The amplification of certain noise artifacts, clearly seen in the result of Fig. \ref{fig:demo}(d), shows that, in addition to learning how to resolve fine details in the artifact, PhENN has learnt, somewhat undesirably, to edge enhance (since all the examples it was trained with were also edge enhanced.) These observations should present fertile ground for further improvements upon the work presented here. 

\section*{Funding}
This research was funded by the Singapore National Research Foundation through the SMART program (Singapore-MIT Alliance for Research and Technology) and by the Intelligence Advanced Research Projects Activity (iARPA) through the RAVEN Program.

\bibliography{sample}

\end{document}